\documentclass[11pt]{article}

\usepackage[preprint]{neurips_2025}

\usepackage[utf8]{inputenc} 
\usepackage[T1]{fontenc}    
\usepackage{natbib} 
\usepackage{url}            
\usepackage{booktabs}       
\usepackage{amsfonts}       
\usepackage{nicefrac}       
\usepackage{xcolor}         
\usepackage{hyperref}
\usepackage{graphicx}
\usepackage{float} 
\usepackage{subcaption} 

\newcommand{\justification}[1]{\textit{#1}}

\title{Pharmacophore-Guided Generative Design of Novel Drug-Like Molecules}

\author{%
    Ekaterina Podplutova$^{1}$ \quad  Anastasia Vepreva$^{1}$ \smallskip \\ \textbf{Olga Konovalova$^{1}$ \quad Vladimir Vinogradov$^{1}$ \quad  Dmitrii Shkil$^{2}$ \quad Andrei Dmitrenko$^{1,3}$}  \medskip \\
$^{1}$Center for AI in Chemistry, ITMO University, St. Petersburg, Russia  \smallskip \\
$^{2}$Moscow Center for Advanced Studies, Moscow, Russia  \smallskip \\
$^{3}$D ONE AG, Zurich, Switzerland  \smallskip \\
 \smallskip \\
\texttt{dmitrenko@pish.itmo.ru}
}

\begin{document}

\maketitle

\begin{abstract}

The integration of artificial intelligence (AI) in early-stage drug discovery offers unprecedented opportunities for exploring chemical space and accelerating hit-to-lead optimization. However, using docking as a reward function during generative model training is computationally expensive and may yield inaccurate results. Here, we present a novel generative framework that balances pharmacophore similarity to reference compounds with structural diversity from active molecules. The framework allows users to provide custom reference sets, including FDA-approved drugs or clinical candidates, and guides the \justification{de novo} generation of potential therapeutics. We demonstrate its applicability through a case study targeting alpha estrogen receptor modulators and antagonists for breast cancer. The generated compounds maintain high pharmacophoric fidelity to known active molecules while introducing substantial structural novelty, suggesting strong potential for functional innovation and patentability. Comprehensive evaluation of the generated molecules against common drug-like properties confirms the robustness and pharmaceutical relevance of the approach.

\end{abstract}

\section{Introduction}

The integration of artificial intelligence (AI) in early-stage drug discovery is transforming pharmaceutical paradigms, enabling more efficient exploration of chemical space and accelerating hit-to-lead progression \cite{ocana2025integrating}. 
Traditional method for accessing biological activity is molecular docking calculation, which predicts the binding affinity between a ligand and its target protein. However, this approach is computationally expensive \cite{Torres2019} when performed iteratively and often yields unreliable scores. Furthermore, it often oversimplifies the complex interactions involved, leading to inaccuracies. Many scoring functions are based on linear energy combinations, which may not adequately capture the nuances of protein-ligand interactions, resulting in poor correlation with experimental binding affinities \cite{zheng2022, yu2023}. 

Pharmacophore-guided methods present a compelling alternative: by focusing on the spatial arrangement of key interaction features (e.g., hydrogen bond donors/acceptors, aromatic or hydrophobic moieties), they provide a more interpretable and robust proxy for biological activity across diverse chemical scaffolds. Pharmacophore-aware similarity measures and latent-space methods have been explored, but few approaches combine pharmacophore-level similarity with structural diversity in seed fragments while explicitly optimizing for docking performance. Existing frameworks like DrugMetric use VAE-based chemical space distances for molecular generation and scaffold diversity \cite{li2024drugmetric, lee2021drug}. Other methods focus on generative modeling of molecular latent spaces (e.g., NP-VAE, conditional $\beta$-VAE), achieving high novelty scores but often sacrificing docking fidelity or pharmacophoric consistency \cite{wu2022cross, gadiya2023pemt, pmc7577280}.

In this work, we present a framework for \justification{de novo} molecule generation that maximizes pharmacophoric similarity to reference compounds (e.g., FDA-approved drugs) while minimizing structural similarity to improve novelty and potential patentability. We demonstrate the utility of this method through a case study targeting estrogen receptor inhibitors for breast cancer. The generated compounds show strong pharmacophoric alignment with known degraders while maintaining high structural diversity. They were further validated using docking scores and synthetic accessibility. The code and data used in this study will be made available at GitHub in the camera-ready version of the manuscript.

\section{Related works}

Recent advances have proposed various frameworks for pharmacophore-aware molecular generation. Zhu et al. introduced PGMG, a graph-based generative model guided by pharmacophoric constraints, which achieved high validity, novelty, and docking scores \cite{zhu2023pgmg}. Seo and Kim developed PharmacoNet, an automated pipeline for pharmacophore model construction and scoring, which accelerates virtual screening while retaining high accuracy \cite{seo2024pharmaconet}. Yu et al. proposed DiffPhore, a diffusion-based model that learns to generate molecules conditioned on pharmacophoric maps and can predict binding poses without explicit docking \cite{yu2025diffphore}. Moyano-Gómez et al. presented O-LAP, which creates cavity-filling pseudo-ligands to improve docking rescoring and account for protein-ligand shape complementarity \cite{moyano-gomez2024olap}. Alakhdar et al. introduced PharmaDiff, a pharmacophore-conditioned diffusion model that generates molecules satisfying 3D feature constraints with improved docking performance \cite{alakhdar2025pharmadiff}.

While existing methods often optimize docking scores or rely on specific binding pockets, our framework is target-agnostic and docking-independent, using pharmacophore similarity as a proxy for biological relevance. Unlike PGMG and PharmaDiff, it balances scaffold novelty with pharmacophoric fidelity; unlike O-LAP and PharmacoNet, it avoids predefined binding sites, enabling early-stage exploration when structural data is lacking. This allows us to access diverse, patentable chemical space while preserving pharmacophoric patterns linked to activity.

\section{Experiments}
\subsection{Overview of the proposed pipeline}
\label{Overview of the proposed pipeline}

\begin{figure}[h!]
    \centering
    \includegraphics[width=1\linewidth]{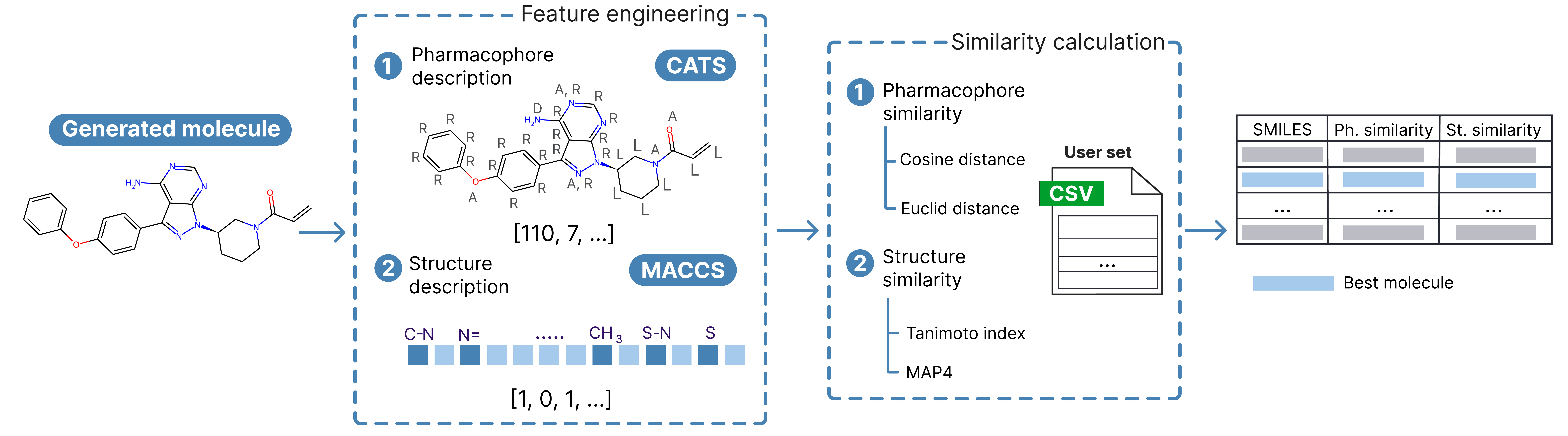}
    \caption{Schematic representation of proposed pipeline.}
    \label{fig:pipeline}
\end{figure}

We present a novel methodology for evaluating the biological activity of molecules that integrates both structural and pharmacophoric similarity assessments against a predefined set of reference compounds (\autoref{fig:pipeline}).

This approach was implemented within the reward function of the reinforcement learning (RL) model, FREED++ \citep{https://doi.org/10.48550/arxiv.2401.09840}. During each cycle of the RL process, generated molecules are encoded using two distinct molecular representations: CATS (Chemically Advanced Template Search) descriptors  \citep{Reutlinger2013}, which capture pharmacophore patterns, and MACCS (Molecular ACCess System) keys \citep{Yang2022}, which represent substructural features.
To compute similarity, the resulting representations are compared to those of the molecules in a user-provided reference set. Given the distinct nature of the two representations, different similarity metrics were employed:

\begin{itemize}
    \item Pharmacophoric similarity, derived from the continuous-valued CATS descriptors, was quantified using cosine similarity and Euclidean distance.
    \item Structural similarity, based on the binary MACCS fingerprints, was assessed using the Tanimoto coefficient, while MAP4 (MinHashed Atom-Pair fingerprint up to four bonds) provides a more expressive representation by combining atom-pair relationships with circular and thus shows higher scores \citep{capecchi2020map4}.
\end{itemize}

The reward function was explicitly designed to simultaneously maximize pharmacophoric similarity and minimize structural similarity to the reference molecules. This dual-objective optimization is critical for generating novel compounds that are likely to retain the desired biological activity (guided by pharmacophore overlap) while exhibiting sufficient structural novelty to enhance their potential for patentability.


\subsection{Baseline evaluation}

As a reference point, we combined QED scoring with docking simulations using QVina. Docking was performed using the crystallographic structure of the alpha-estrogen receptor (PDB ID: 8AWG). The target was selected due to its central role in breast cancer pathogenesis and the availability of a high-resolution validated structure.

\subsection{Reward function variants}

As detailed in \autoref{Overview of the proposed pipeline}, pharmacophore similarity was evaluated using cosine and Euclidean distances. Cosine similarity evaluates the orientation of vectors and is widely used for molecular fingerprints, while Euclidean distance captures both magnitude and direction, providing a complementary measure of dissimilarity. Structural similarity was assessed using the Tanimoto coefficient and MAP4. We tested four configurations of our reward function:

\begin{enumerate}
    \item QED + Tanimoto  + Euclidean similarity 
    \item QED + Tanimoto + Cosine similarity
    \item QED + MAP4 + Euclidean similarity
    \item QED + MAP4 + Cosine similarity
\end{enumerate}


\subsection{Additional profiling}

Generated molecules were further evaluated with orthogonal filters. Synthetic accessibility (SA) scores estimated practical feasibility, and novelty was quantified by checking absence from ChEMBL, ZINC, and PubChem databases.

Finally, we analyzed the distributions of QED, docking scores, and molecular properties including SA, MAP4, Tanimoto, and pharmacophore similarity assessed via Euclidean and Cosine metrics \autoref{Distribution of key properties evaluated in experiments}. 

\section{Results and Discussion}

\subsection{Overall Pharmacophore and Drug-Likeness Assessment}

 The evaluation of generated molecules across different reward configurations highlights the framework's ability to optimize both pharmacophoric similarity and predicted binding affinity (Table~\ref{table_results}). The baseline molecules, generated without pharmacophore rewards, show relatively good predicted binding affinity (docking score of -8.65), complete novelty (100\%), but low drug-likeness (QED of 0.30). Despite achieving more favorable docking scores, the baseline generated molecules display very low pharmacophoric similarity to established drugs, raising concerns about their biological relevance. Additionally, their synthetic accessibility remains in question (SA score of 6.28).

\begin{table}[h!]
\centering
\caption{Evaluation of generated molecules across different reward configurations (mean $\pm$ std).}
\label{table_results}
\resizebox{\textwidth}{!}{%
\renewcommand{\arraystretch}{1.2}
\begin{tabular}{lccccccccc}
\hline
\multicolumn{1}{c}{Setup} &
  \begin{tabular}[c]{@{}c@{}}Tanimoto\\ index (↓)\end{tabular} &
  \begin{tabular}[c]{@{}c@{}}MAP4\\ score (↓)\end{tabular} &
  \begin{tabular}[c]{@{}c@{}}Cosine\\ similarity (↑)\end{tabular} &
  \begin{tabular}[c]{@{}c@{}}Euclid\\ similarity (↓)\end{tabular} &
  QED (↑) &
  \begin{tabular}[c]{@{}c@{}}Docking\\ score (↓)\end{tabular} &
  \begin{tabular}[c]{@{}c@{}}SA\\ score (↓)\end{tabular} &
  Novelty (↑)\\ \hline

Baseline & \textbf{0.34 $\pm$ 0.05} & \textbf{0.03 $\pm$ 0.01} & 0.58 $\pm$ 0.27 & 70.3 $\pm$ 13.03 & 0.30 $\pm$ 0.08 & \textbf{-8.64 $\pm$ 1.03} & 6.28 $\pm$ 0.64 & \textbf{100} \\ \hline

Setup 1 & \textbf{0.34 $\pm$ 0.05} & 0.04 $\pm$ 0.01 & \textbf{0.94 $\pm$ 0.06} & \textbf{34.80 $\pm$ 7.84} & 0.33 $\pm$ 0.13 & -6.49 $\pm$ 1.17 & 4.64 $\pm$ 0.51 & \textbf{100} \\

Setup 2 & 0.36 $\pm$ 0.05 & \textbf{0.03 $\pm$ 0.01} & 0.83 $\pm$ 0.05 & 54.92 $\pm$ 8.60 & \textbf{0.59 $\pm$ 0.16} & -6.71 $\pm$ 0.55 & 4.72 $\pm$ 0.49 & 99.6 \\

Setup 3 & 0.35 $\pm$ 0.05 & 0.04 $\pm$ 0.01 & \textbf{0.94 $\pm$ 0.06} & 50.47 $\pm$ 10.16 & 0.44 $\pm$ 0.16 & -7.09 $\pm$ 0.66 & 4.67 $\pm$ 0.45 & 84.5 \\

Setup 4 & 0.35 $\pm$ 0.05 & \textbf{0.03 $\pm$ 0.01} & 0.87 $\pm$ 0.07 & 38.92 $\pm$ 9.37 & 0.34 $\pm$ 0.15 & -6.47 $\pm$ 1.02 & \textbf{4.61 $\pm$ 0.50} & \textbf{100} \\ \hline

\end{tabular}%
}
\end{table}

\begin{figure}[H]
    \centering

    \begin{subfigure}{0.45\textwidth}
        \includegraphics[width=\linewidth]{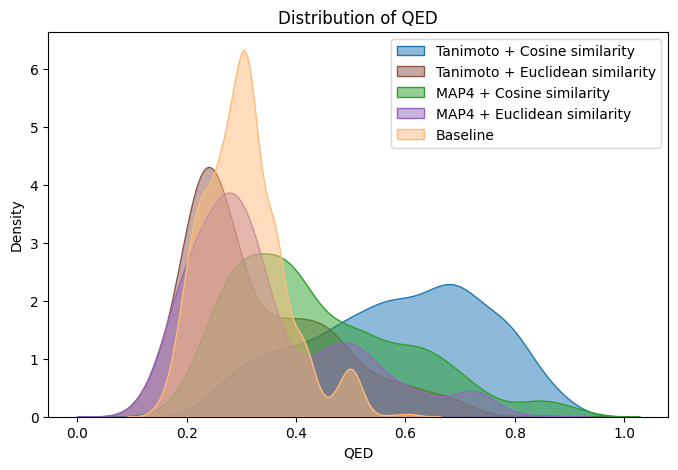}
        \caption{QED}
        \label{fig:qed}
    \end{subfigure}
    \hfill
    \begin{subfigure}{0.45\textwidth}
        \includegraphics[width=\linewidth]{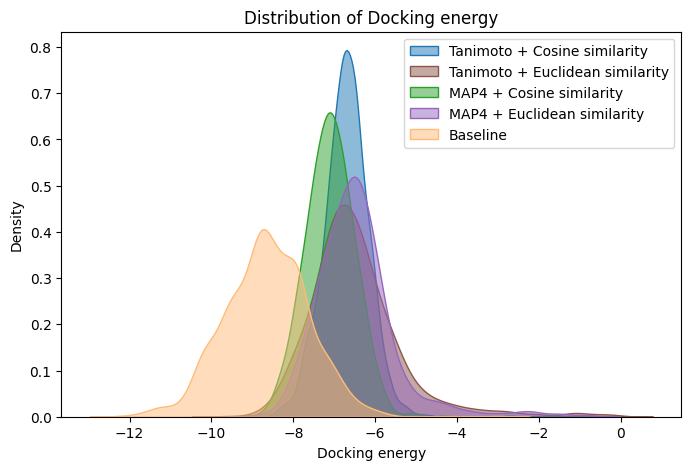}
        \caption{Docking energy}
        \label{fig:docking}
    \end{subfigure}

    \begin{subfigure}{0.45\textwidth}
        \includegraphics[width=\linewidth]{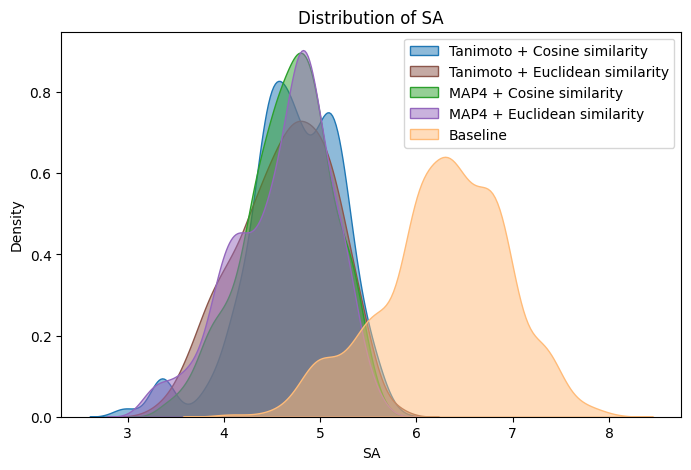}
        \caption{SA}
        \label{fig:sa}
    \end{subfigure}
    \hfill
    \begin{subfigure}{0.45\textwidth}
        \includegraphics[width=\linewidth]{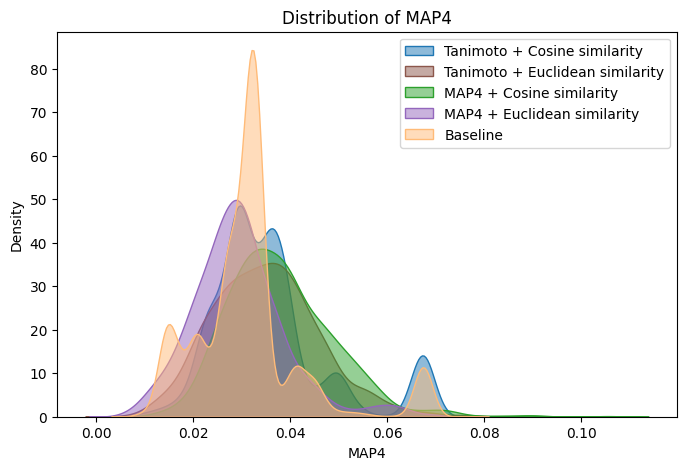}
        \caption{MAP4}
        \label{fig:map4}
    \end{subfigure}

    \begin{subfigure}{0.45\textwidth}
        \includegraphics[width=\linewidth]{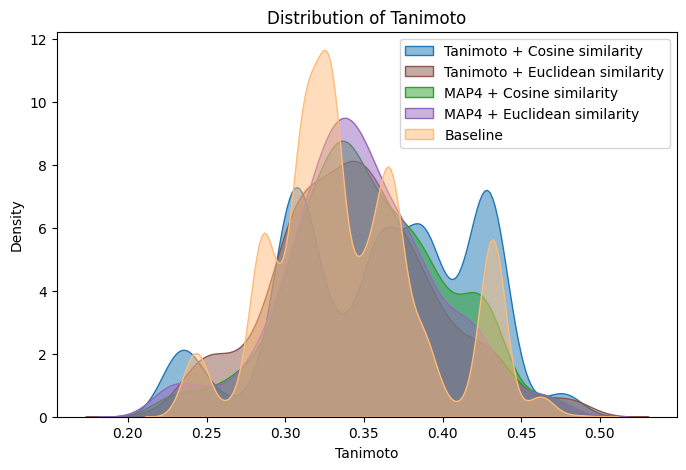}
        \caption{Tanimoto}
        \label{fig:tanimoto}
    \end{subfigure}
    \hfill
    \begin{subfigure}{0.45\textwidth}
        \includegraphics[width=\linewidth]{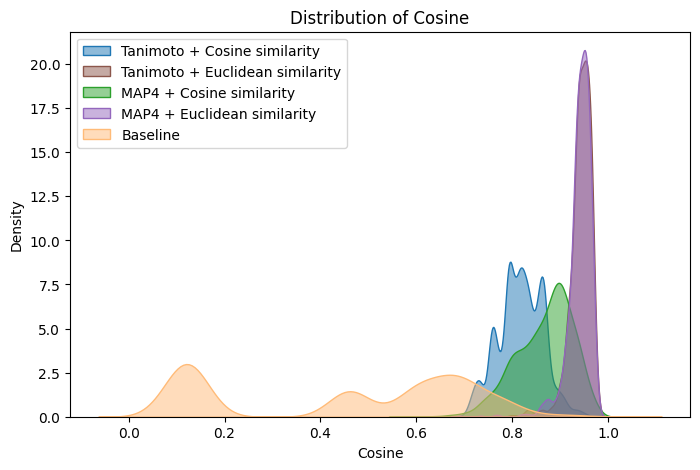}
        \caption{Cosine}
        \label{fig:cosine}
    \end{subfigure}

    \begin{subfigure}{0.45\textwidth}
        \includegraphics[width=\linewidth]{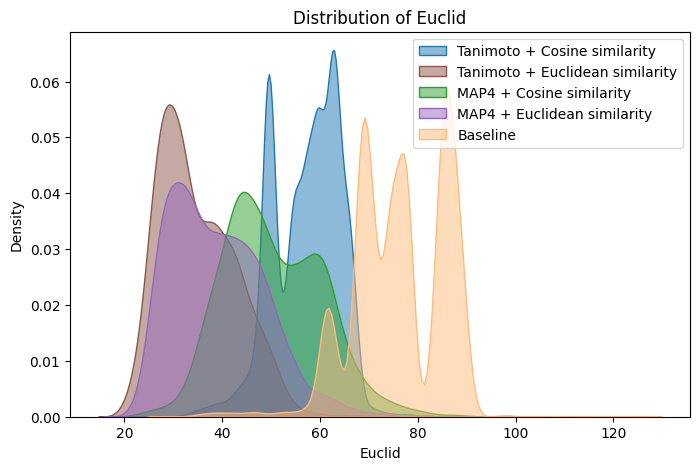}
        \caption{Euclid}
        \label{fig:euclid}
    \end{subfigure}

    \caption{Distributions of key properties evaluated in experiments.}
    \label{fig:distributions}
\end{figure}

Introducing pharmacophore similarity and structural diversity in reward functions (Setups 1-4) led to improved molecular properties, with QED values and SA scores improving across pharmacophore-guided setups. This suggests that enforcing pharmacophoric fidelity encourages the generation of more drug-like and synthetically accessible molecules. The impact of different similarity metrics on these property profiles is visually assessed on Figure 2. Specifically, the QED distribution (Figure \ref{fig:qed}) for the baseline is concentrated around 0.3-0.4, while MAP4 + Cosine similarity shifts this distribution towards higher values (peak near 0.6-0.7), indicating improved drug-likeness. Similarly, the SA distribution (Figure \ref{fig:sa}) shows a lower peak for the other methods in comparison to the baseline which has a peak at 4, suggesting improved synthetic accessibility. The docking score distribution (Figure \ref{fig:docking}) is shifted towards less negative values for all setups compared to the baseline (peak around -8), indicating lower binding affinity. However, the average docking score of the known alpha-estrogen receptor modulators and antagonists, which served as the basis for the pharmacophore descriptors, was -6.64. This allows us to conclude that all four proposed setups are, in fact, comparable to the confirmed receptor modulators and antagonists in binding affinity, assessed by the docking score. Furthermore, cosine similarity (Figure \ref{fig:cosine}) is higher for MAP4 + Cosine similarity compared to Tanimoto + Cosine similarity which has a peak near 0.7, indicating that the MAP4 + Cosine similarity method generates structures with a higher average cosine similarity score.

MAP4 provides a rich molecular representation, encoding atom-pair relationships and leveraging MinHash to capture global topology and local motifs efficiently. Pharmacophoric and structural similarity values remain comparable across all reward setups, showing that our framework generates molecules with favorable predicted binding affinity, drug-likeness, and structural novelty.

\begin{figure}[H]
    \centering
    \includegraphics[width=0.9\linewidth]{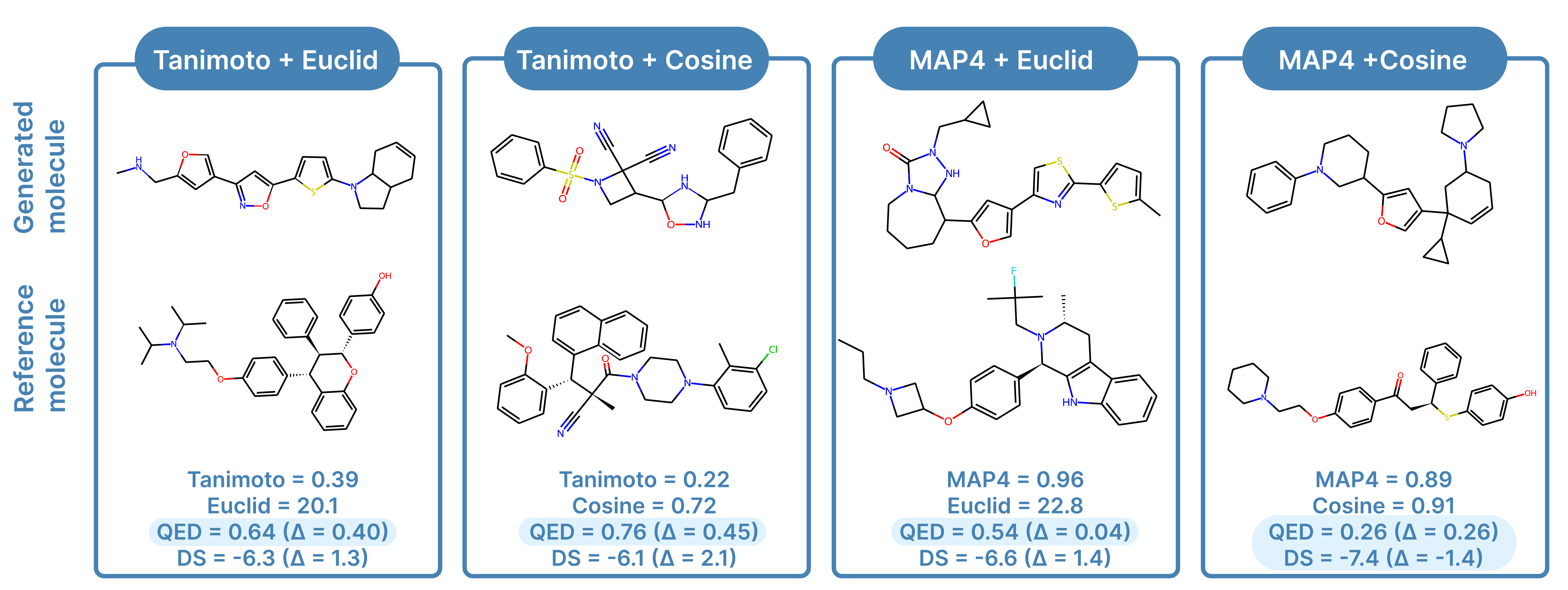}
    \caption{Best generated molecules and their pharmacophore analogue.}
    \label{fig:best_molecules}
\end{figure}

In \autoref{fig:best_molecules}, representative generated molecules and their reference analogs (one per reward setting) reproduce key pharmacophoric patterns, tri-aromatic/heteroaromatic motifs with similar linker lengths, while reshaping scaffolds. Even though docking score improvement is notable mainly in the MAP4 + cosine setup, the top molecules exhibit higher QED than reference degraders.

These results indicate that our reward functions drive convergence on biologically meaningful pharmacophoric arrangements (aromatic triads, conserved H-bond vectors, hydrophobic spacers) without collapsing to close structural analogs, balancing functional similarity and scaffold novelty.


\subsection{Methodological limitations}

This study is subject to several methodological limitations. While the generated molecules exhibit high pharmacophoric similarity to known degraders, they demonstrate only moderate docking scores and QED. The present approach, which employs a constrained set of pharmacophore descriptors and similarity metrics, may inherently limit the diversity of the generated molecular scaffolds. It is important to emphasize that this work does not propose the replacement of docking simulations; rather, it suggests their application at a subsequent stage for filtering outputs, as opposed to their integration into the generative reward function. Future research will focus on extending this framework through the incorporation of alternative pharmacophore representations, additional similarity measures, and more diverse generative models to concurrently enhance biological relevance and chemical novelty. As a preliminary investigation, this work establishes a foundation for significant further development and refinement, culminating in experimental validation via synthesis and biological assays.

\section{Conclusion and Future work}

We proposed a pharmacophore-guided generative approach for designing potentially active and selective molecules using a reinforcement learning model. Pharmacophoric similarity was evaluated with CATS descriptors using Euclidean and cosine metrics, while structural novelty was encouraged by minimizing similarity based on MACCS descriptors using the classical Tanimoto coefficient, as well as the recently proposed MAP4 metric. In a case study targeting estrogen receptor inhibitors for breast cancer, the generated compounds showed high pharmacophoric similarity to known actives and low structural similarity, suggesting strong novelty and patentability. All molecules also met basic drug-like criteria, supporting the method’s potential for further development and experimental validation.

\section{Acknowledgment}

This work supported by the Ministry of Economic Development of the Russian Federation (IGK 000000C313925P4C0002), agreement No139-15-2025-010.

\bibliographystyle{unsrt}
\bibliography{references}

\section{Technical Appendices and Supplementary Material}

\subsection{Distribution of key properties evaluated in experiments}
\label{Distribution of key properties evaluated in experiments}

\subsection{Best generated molecules}

\section*{NeurIPS Paper Checklist}

\begin{enumerate}

\item {\bf Claims}
    \item[] Question: Do the main claims made in the abstract and introduction accurately reflect the paper's contributions and scope?
    \item[] Answer: \textbf{Yes} 
    \item[] Justification: \justification{The abstract and introduction consistently highlight the key methodological contribution pharmacophore-guided generative design and frame it within the context of existing challenges in drug discovery.}

\item {\bf Limitations}
    \item[] Question: Does the paper discuss the limitations of the work performed by the authors?
    \item[] Answer: \textbf{Yes} 
    \item[] Justification: \justification{The limitations section explicitly acknowledges methodological constraints, including reliance on docking as a proxy for biological activity, the potential bias introduced by pharmacophore similarity metrics, and the restricted scope imposed by reference sets. The absence of wet-lab validation is also stated, reinforcing transparency.}

\item {\bf Theory assumptions and proofs}
    \item[] Question: For each theoretical result, does the paper provide the full set of assumptions and a complete (and correct) proof?
    \item[] Answer: \textbf{N/A} 
    \item[] Justification: \justification{The paper does not include theoretical results.}

    \item {\bf Experimental result reproducibility}
    \item[] Question: Does the paper fully disclose all the information needed to reproduce the main experimental results of the paper to the extent that it affects the main claims and/or conclusions of the paper (regardless of whether the code and data are provided or not)?
    \item[] Answer: \textbf{Yes} 
    \item[] Justification: \justification{The paper provides complete disclosure of experimental settings, including dataset composition, choice of reference molecules, pharmacophore descriptors, similarity metrics, model architecture, and training parameters. These details are sufficient to reproduce the reported results and support the main conclusions, independent of access to code or supplementary material.}

\item {\bf Open access to data and code}
    \item[] Question: Does the paper provide open access to the data and code, with sufficient instructions to faithfully reproduce the main experimental results, as described in supplemental material?
    \item[] Answer: \textbf{Yes} 
    \item[] Justification: \justification{The authors release both the trained models and generation pipeline with detailed usage instructions. Data splits, input reference molecules, and evaluation metrics are included in the GitHub.}

\item {\bf Experimental setting/details}
    \item[] Question: Does the paper specify all the training and test details (e.g., data splits, hyperparameters, how they were chosen, type of optimizer, etc.) necessary to understand the results?
    \item[] Answer: \textbf{Yes} 
    \item[] Justification: \justification{Training and evaluation settings are reported in GitHub.}

\item {\bf Experiment statistical significance}
    \item[] Question: Does the paper report error bars suitably and correctly defined or other appropriate information about the statistical significance of the experiments?
    \item[] Answer: \textbf{Yes} 
    \item[] Justification: \justification{The results for all experimental setups are now accompanied by standard deviations for each metric, providing clear quantitative measures of variability.}

\item {\bf Experiments compute resources}
    \item[] Question: For each experiment, does the paper provide sufficient information on the computer resources (type of compute workers, memory, time of execution) needed to reproduce the experiments?
    \item[] Answer: \textbf{Yes} 
    \item[] Justification: \justification{Calculations were performed on a server with an NVIDIA A6000 GPU (20 GB RAM).}
    
\item {\bf Code of ethics}
    \item[] Question: Does the research conducted in the paper conform, in every respect, with the NeurIPS Code of Ethics \url{https://neurips.cc/public/EthicsGuidelines}?
    \item[] Answer: \textbf{Yes} 
    \item[] Justification: \justification{The study complies with ethical standards.}

\item {\bf Broader impacts}
    \item[] Question: Does the paper discuss both potential positive societal impacts and negative societal impacts of the work performed?
    \item[] Answer: \textbf{Yes} 
    \item[] Justification: \justification{The discussion covers positive impacts, such as accelerating drug discovery and enabling patentable molecule design, as well as risks, including potential misuse of generative methods to design harmful compounds.}
    
\item {\bf Safeguards}
    \item[] Question: Does the paper describe safeguards that have been put in place for responsible release of data or models that have a high risk for misuse (e.g., pretrained language models, image generators, or scraped datasets)?
    \item[] Answer: \textbf{N/A} 
    \item[] Justification: \justification{The paper poses no such risks.}

\item {\bf Licenses for existing assets}
    \item[] Question: Are the creators or original owners of assets (e.g., code, data, models), used in the paper, properly credited and are the license and terms of use explicitly mentioned and properly respected?
    \item[] Answer: \textbf{Yes} 
    \item[] Justification: \justification{The paper clearly credits the sources of all external assets, including molecular datasets, similarity metrics and filtering tools. References to the original publications are provided, and the use of these resources complies with their respective licenses and terms of use.}

\item {\bf New assets}
    \item[] Question: Are new assets introduced in the paper well documented and is the documentation provided alongside the assets?
    \item[] Answer: \textbf{Yes} 
    \item[] Justification: \justification{The newly introduced assets are described in detail within the paper and supplementary material. Clear documentation and usage instructions are provided alongside the released code repository, ensuring that other researchers can easily understand, reproduce, and extend the work.}

\item {\bf Crowdsourcing and research with human subjects}
    \item[] Question: For crowdsourcing experiments and research with human subjects, does the paper include the full text of instructions given to participants and screenshots, if applicable, as well as details about compensation (if any)? 
    \item[] Answer: \textbf{N/A} 
    \item[] Justification: \justification{The paper does not involve crowdsourcing nor research with human subjects.}

\item {\bf Institutional review board (IRB) approvals or equivalent for research with human subjects}
    \item[] Question: Does the paper describe potential risks incurred by study participants, whether such risks were disclosed to the subjects, and whether Institutional Review Board (IRB) approvals (or an equivalent approval/review based on the requirements of your country or institution) were obtained?
    \item[] Answer: \textbf{N/A} 
    \item[] Justification: \justification{The paper does not involve crowdsourcing nor research with human subjects.}

\item {\bf Declaration of LLM usage}
    \item[] Question: Does the paper describe the usage of LLMs if it is an important, original, or non-standard component of the core methods in this research? Note that if the LLM is used only for writing, editing, or formatting purposes and does not impact the core methodology, scientific rigorousness, or originality of the research, declaration is not required.
    \item[] Answer: \textbf{N/A} 
    \item[] Justification: \justification{The core method development in this research does not involve LLMs as any important, original, or non-standard components.}

\end{enumerate}

\end{document}